\documentclass{llncs}
\usepackage[T1]{fontenc}
\usepackage{graphicx}
\usepackage{comment}
\usepackage{booktabs}
\usepackage{tabularx}
\usepackage{multirow}

\newcommand{\mycomment}[1]{} 

\begin{document}
\title{Commit-Aware Learning-Based Test Case Prioritization for Continuous Integration}
%
%
\author{Lorenzo Abbondante\orcidID{0009-0007-6561-1370} 
Gerardo Canfora\orcidID{0000-0003-0049-1279} }
\authorrunning{L. Abbondante, G. Canfora}
%
\institute{University of Sannio, Dept. of Engineering, Benevento - Italy}
\maketitle              
\begin{abstract}
Regression testing in Continuous Integration (CI) pipelines is increasingly costly due to the growing size and execution frequency of test suites. Test Case Prioritization (TCP) mitigates this problem by reordering tests to expose faults earlier. However, most existing techniques rely primarily on historical execution data and coverage metrics, neglecting the rich structural information contained in code changes.

We propose a commit-aware, learning-based TCP method that combines structural properties of version-control diffs, test coverage relations, and historical execution behavior into a unified predictive model. Given a new commit, the method estimates the probability that each test suite will reveal at least one failure and prioritizes test execution accordingly.

We evaluate our method on five Defects4J projects using a leave-one-project-out cross-project validation setting. Results show that the commit-aware TCP significantly outperform non–commit-aware baselines in both classification and prioritization effectiveness. 

Our findings show that including commit structural semantics substantially enhances regression fault detection and enables robust, generalizable learning-based TCP in CI environments.

\end{abstract}

\section{Introduction}
\label{sec:intro}

In modern Agile and DevOps environments, software systems evolve continuously through rapid and incremental development cycles. Continuous Integration (CI) and Continuous Deployment (CD) pipelines automate the process of building, testing, and deploying new code as soon as it is committed to the shared repository. Although these practices significantly improve responsiveness and shorten delivery times, they also increase the frequency with which large regression test suites must be executed.
In large-scale industrial projects, CI pipelines may execute thousands of test cases multiple times per day under strict time and resource constraints. Executing the full test suite after every commit becomes prohibitively expensive, and organizations face the need for intelligent strategies that optimize testing efficiency without compromising the ability to detect regressions early. Test Case Prioritization (TCP) addresses this challenge by reordering test execution so that test cases with the highest probability of revealing faults are executed earlier~\cite{DBLP:journals/tse/Rothermel,DBLP:journals/stvr/YooH12}.

Over the past two decades, many researchers have investigated TCP strategies. Early approaches relied primarily on static heuristics and structural coverage data~\cite{DBLP:journals/tse/ElbaumMR02,DBLP:conf/icse/GravesHKPR98,DBLP:journals/tse/Rothermel}.
Subsequent work introduced \textit{search-based} and \textit{metaheuristic} optimizations—such as Genetic Algorithms—that treat TCP as a multi-objective problem balancing fault detection, coverage, and cost~\cite{DBLP:journals/tse/LiHH07,singhal2023aco}.

More recently, the advent of AI and deep learning has opened new possibilities for data-driven TCP, where models learn to predict the fault-revealing potential of tests based on historical and structural features~\cite{DBLP:journals/tse/BagherzadehKB22,DBLP:journals/corr/abs-2206-15428TEST2VEC,DBLP:conf/icsm/SharifML21,DBLP:conf/icsm/Zhao0023}. These AI-based methods outperform many traditional heuristics but often rely heavily on historical execution data, overlooking the contextual information contained in the code changes that trigger each CI cycle.

Indeed, although version-control repositories encode rich signals about code evolution, very few TCP techniques leverage the semantics of commits. Every commit conveys structural change information—such as the number of altered files, the added or removed lines, and the churn level—that has long been recognized in defect prediction research as a strong predictor of fault-proneness. However, this knowledge has been underutilized in prioritizing the tests executed after each change.

We address this gap by introducing a \textit{commit-aware, machine-learning-based TCP method} that integrates version control diffs, test coverage information, and historical outcomes into a unified predictive model. We address the hypothesis that commit structural characteristics combined with coverage and execution history provide stronger signals for test failure prediction than historical data alone.  The intuition is that tests that exercise recently modified or high-churn components are more likely to detect regressions. By learning the relationships between code evolution and test failures, the proposed approach bridges the traditional ``black-box'' (based on execution history) and ``white-box'' (based on coverage measures) paradigms into a strategy suitable for CI environments. In addition to improving responsiveness to current changes, commit-aware TCP helps mitigate cold-start and data sparsity. In fact, history-based methods degrade when many tests have never failed or when the history is short or noisy. The features related to the current diff provide useful signals even with a limited history.

The remainder of this paper is structured as follows.
Section~\ref{sec:literature} discusses related work.  
Section~\ref{sec:method} introduces the proposed commit-aware method, while 
Section~\ref{sec:setup} describes the experimental methodology.  
Section~\ref{sec:results} reports and analyzes the  results, and 
Section~\ref{sec:conclusion} concludes the paper.

\section{Related work}
\label{sec:literature}

TCP has been extensively studied in the context of regression testing and CI, with the goal of accelerating fault detection by reordering the execution of the tests~\cite{DBLP:journals/tse/Rothermel,DBLP:journals/stvr/YooH12}. The majority of prior work can be grouped into three main paradigms: traditional heuristic-based approaches, search-based/metaheuristic optimization, and learning-based techniques.

\subsection{Traditional and Search-Based TCP}
Early TCP research focused on deterministic heuristics based on structural coverage and execution cost~\cite{DBLP:journals/tse/Rothermel,DBLP:journals/tse/ElbaumMR02,DBLP:conf/icse/GravesHKPR98}. Whilst these approaches demonstrated that coverage-driven ordering can significantly accelerate fault detection, they often fail to adapt to evolving code bases and CI dynamics.
Subsequent work formulated TCP as an optimization problem, applying metaheuristics such as Genetic Algorithms, Particle Swarm Optimization, and Firefly algorithms to balance multiple objectives including fault detection, execution cost, and severity~\cite{DBLP:journals/tse/LiHH07,Ashraf2017,9103716,DBLP:journals/access/KhatibsyarbiniI19}. While effective in controlled settings, these methods are scarcely scalable and require expensive re-optimization when project characteristics change.

Efforts to improve traditional deterministic methods include multi-objective scoring systems that balance effectiveness and resource constraints~\cite{DBLP:journals/sp/SamadMKIB21,Samad2021}, as well as the Accelerated Greedy Additional algorithm, which specifically targets the efficiency bottleneck of greedy approaches by reducing computational complexity~\cite{DBLP:journals/tse/LiZLHZ22}.

Information-Retrieval and similarity-based techniques introduce lightweight alternatives that prioritize test diversity and change relevance~\cite{DBLP:conf/icse/MirandaCVB18,DBLP:conf/icst/HemmatiFM15,DBLP:conf/issta/PengSZ20}. History-based and risk-driven models further exploit execution logs, commit metadata, and defect-prediction metrics to guide prioritization~\cite{DBLP:conf/icst/HemmatiFM15,DBLP:conf/icst/PatersonCAKFM19,DBLP:conf/icse/LiangER18,DBLP:journals/ase/MahdiehMM22,Ahmed,DBLP:journals/tosem/ChenCWZWCZW23}. Large-scale industrial studies show that although sophisticated heuristics can be beneficial, simple history- and time-based strategies remain surprisingly competitive in practice~\cite{DBLP:conf/icsm/LuoMPP18,DBLP:conf/issta/ChengWJM24Revisiting,DBLP:journals/ase/WangYWDZCW25}.

\subsection{Learning-Based TCP}
Learning-based TCP methods treat prioritization as a predictive ranking problem, learning from historical execution data, coverage, and code metrics~\cite{DBLP:journals/ese/PanBGB22,DBLP:journals/access/KhatibsyarbiniI21}. Early industrial approaches applied supervised classifiers to predict failing tests based on execution history and change-aware features~\cite{DBLP:conf/sigsoft/BusjaegerX16,DBLP:conf/icse/MachalicaSP019,DBLP:journals/ese/HajriGPB20,DBLP:journals/sqj/Marijan23}, while later work employed ensemble models and automated hyperparameter tuning to improve robustness~\cite{DBLP:conf/icst/KhanALSCST24}. While some authors focus on model training, others define tools to collect the features evaluated in their approach~\cite{DBLP:journals/tse/YaraghiBKB23}.

More recent research explores reinforcement learning and learning-to-rank formulations, enabling adaptive prioritization under CI feedback loops~\cite{DBLP:journals/tse/BagherzadehKB22,DBLP:conf/issta/SpiekerGMM18,DBLP:conf/icse/BertolinoGMPR20,DBLP:journals/infsof/RozaLV24,DBLP:journals/access/AlrakbanAA25,DBLP:journals/infsof/QianYZLF25}. Deep learning approaches further model long-term execution patterns and sequential dependencies using neural and recurrent architectures~\cite{DBLP:conf/icsm/SharifML21,BEHERA20254070,DBLP:conf/aitest/AbdelkarimE23,DBLP:conf/qrs/JabbarHF23,DBLP:conf/wcre/RozaLSV22,DBLP:journals/access/VescanT25,DBLP:conf/kbse/VescanGS23,DBLP:journals/infsof/MahdiehMENJ20}. Transfer learning and clustering-based strategies have been proposed to mitigate data scarcity and volatility across projects~\cite{DBLP:conf/ast/MamataALSCST23,DBLP:conf/icsm/VescanS20,DBLP:journals/infsof/WangZ24,DBLP:journals/corr/abs-2404-16395}, while others have framed TCP as a Learning-to-Rank task using SVMRank on natural-language descriptions~\cite{DBLP:conf/icmla/LachmannSNSS16}.

Despite these advances, most learning-based approaches rely predominantly on execution history and coverage signals, while structural commit semantics remain weakly exploited, particularly in cross-project generalization settings. The method proposed in this paper directly addresses this limitation by embedding commit-level diff semantics into a unified learning framework and evaluating it under cross-project validation.

\section{The method}
\label{sec:method}

This section introduces our commit-aware, learning-based TCP method. The goal is to predict, for each test suite and each commit, the probability of observing at least one failure and to prioritize tests accordingly.
Whilst traditional TCP focuses on historical data or code coverage, our approach comprises structural code change information extracted from version-control diffs.
The underlying hypothesis is that tests exercising recently modified or high-churn
components are more likely to expose regressions, and that commit-level change
properties provide predictive signals complementary to historical behavior.

Given a commit $c_t$, the framework builds a feature vector for each test suite that encodes:
(i) historical execution behavior,
(ii) structural coverage relations between tests and production code, and
(iii) commit-level change relevance signals.
A supervised machine-learning model estimates the probability that a given test suite will fail under $c_t$,
and test suites are ranked in descending order of predicted failure probability.

\subsection{Design Rationale}

The proposed method combines black-box historical signals with white-box structural information derived from coverage and code evolution.
Commit-level metrics capture change magnitude and locality,
which are well-known correlates of fault-proneness in defect prediction research.

Historical execution features encode temporal behavioral tendencies of test suites,
while coverage features approximate their structural exposure to production code.
By combining these three dimensions, the model can reason about both what changed in the system
and which tests are structurally and behaviorally most affected by the change.

This hybrid representation is particularly suitable for CI environments,
where historical data may be sparse or noisy and purely history-based methods suffer from cold-start effects.
Commit-aware features provide informative signals even when execution history is limited.

\subsection{Feature Representation}

Each test suite under a given commit is represented by a compact set of features.
All features are computed using only information available prior to executing the current test run,
ensuring leakage-free learning.
The complete formal definition of features and the process to compute them is provided in the replication package~\cite{ourReplicationPackage}.

\subsubsection{Commit-Aware (Diff-Based) Features}

Commit-aware features capture the structural properties of code changes introduced by the current commit.
From the version-control diff, we extract basic change metrics such as:
the number of modified files, added lines, and removed lines.

These metrics are originally defined at file or class level.
To obtain test-level representations, we associate each test suite with the production classes it covers, using a heuristic mapping~\cite{ourReplicationPackage},
and propagate diff information from modified classes to their corresponding test suites.
If a test suite covers multiple modified components, diff metrics are aggregated.

The resulting features approximate the structural relevance of each test suite with respect to the current change,
capturing both the size of the change and its proximity to the code exercised by the test.

\subsubsection{Historical Features}

These features summarize the past behavior of test suites.
We consider:
(i) the previous pass rate, representing the most recent execution outcome,
and
(ii) a weighted historical pass rate, computed over multiple past versions.

The weighted pass rate combines temporal recency with commit similarity,
assigning higher importance to recent executions and to past commits with similar change intensity.
This feature estimates the expected reliability of a test suite under the current commit.

\subsubsection{Coverage Features}

Coverage features describe the structural relationship between test suites and production code.
We compute:
(i) the suite coverage ratio, indicating how much of the system is exercised by the test suite,
and
(ii) a combined coverage--diff signal, capturing the extent to which a test suite covers recently modified components.

This combination allows the model to identify test suites that are both structurally broad
and directly exposed to the latest code changes.

\subsection{Machine Learning Models}

We evaluate two representative supervised learning models with complementary characteristics:
gradient-boosted decision trees (XGBoost) and a feedforward neural network (MLP).
XGBoost is well suited for imbalanced tabular data and provides interpretable feature importance,
while MLPs can capture non-linear interactions among heterogeneous features.
Both models are trained to estimate the probability that a test suite will fail under a given commit.

\subsection{Training and Test Prioritization}

Models are trained using a leave-one-project-out strategy to evaluate cross-project generalization. Details are provided in the replication package~\cite{ourReplicationPackage}.
At prediction time, each model outputs a failure probability for each test suite associated with the current commit.
Test suites are then prioritized in descending order of predicted failure probability, producing the final execution order evaluated in the experimental phase.
\section{Experimental setup}
\label{sec:setup}

This section describes the experimental methodology adopted to evaluate the proposed commit-aware TCP approach.
We report research questions, dataset and data processing strategies, evaluation metrics, design and analyses.
Our automated data collection pipeline, detailed implementation artifacts and scripts are provided in the publicly available replication package~\cite{ourReplicationPackage}.

\subsection{Research Questions}

We address the following research questions:

\begin{itemize}
    \item \textbf{RQ1:} Can commit-aware features improve test failure prediction?
    \item \textbf{RQ2:} Does improved failure prediction translate into more effective test case prioritization?
    \item \textbf{RQ3:} How do different supervised learning models compare when using the same commit-aware feature set?
\end{itemize}

RQ1 evaluates the predictive contribution of commit-aware features using classification metrics.
RQ2 assesses whether improved prediction leads to faster and more effective fault detection.
RQ3 compares the behavior of different learning models under identical feature representations,
allowing us to isolate model-specific effects from feature-driven improvements.

\subsection{Metrics}

To answer \textbf{RQ1}, we evaluate failure prediction performance using Precision--Recall AUC (PR-AUC) and F1-score.
These metrics are appropriate for highly imbalanced classification problems,
as they emphasize performance on the minority (failing) class.

To answer \textbf{RQ2}, we evaluate test case prioritization effectiveness using:
(i) Average Percentage of Faults Detected (APFD), that measures the average rate of fault detection per percentage of test suite execution~\cite{DBLP:journals/tse/ElbaumMR02}, and;
(ii) Speedup, measuring the reduction in First-Time-To-Failure relative to chronological execution.

\textbf{RQ3} is addressed by comparing learning models using the same feature set
across both prediction (PR-AUC, F1) and prioritization (APFD, Speedup) metrics.

\subsection{Dataset}

We evaluate the proposed approach on five projects from the Defects4J benchmark~\cite{DBLP:conf/issta/JustJE14}, namely \textit{Math}, \textit{Lang}, \textit{Jsoup}, \textit{Time}, \textit{JacksonDatabind}.
Defects4J provides real-world Java faults, reproducible regression testing scenarios,
and versioned project histories, and is widely used in regression testing research.

Each project consists of multiple faulty program versions,
associated test suites, and execution outcomes.

\subsubsection{Heuristic mapping}
Commit-aware features are originally defined at file or class level, while prediction is performed at test-suite level.
To bridge this gap, we apply a heuristic mapping, based on similarity of file names, that associates each test suite with the production classes it covers.
Diff-based metrics computed on modified classes are then propagated to the corresponding test suites and aggregated when multiple classes are involved.

This mapping enables the estimation of change relevance at test level while remaining scalable and lightweight.
While simple, the heuristic has proven sufficiently robust: we manually checked a sample of 415 test-class mappings taken from the Math project and observed that 89.16\% of test suites are linked to at least one Java class.

\subsubsection{Dataset imbalance}

The dataset exhibits strong class imbalance, as failing test suites represent
a small fraction of the total observations.
To mitigate this issue, we apply standard resampling techniques during training.

Specifically, we applied SMOTE oversampling and random undersampling for the NN-based model, and the \texttt{scale\_pos\_weight} parameter tuning for the decision-tree-based model.
These strategies prevent learning models from being biased toward the majority class
and improve sensitivity to failure instances.

\subsubsection{Data imputation}

In some cases, the heuristic mapping was unable to associate the code changes with the test suites for each commit. To address missing diff-metric values arising from imperfect mappings between production classes and test suites, we applied a data imputation strategy that leverages an Exponentially Weighted Moving Average mechanism with local and global fallbacks to replace null values with statistically coherent estimates. This ensured that the learning models received consistent feature vectors and preserved the structural integrity of positive examples without exacerbating the existing class imbalance.

\subsection{Design}
We use a leave-one-out cross-project strategy: five different clusters each with four training projects and a test project. In the evaluation of the results, we consider only four out of the five folds. The fold where the \textit{Time} project represents the test set is kept out of the evaluation as this project, due to its structure, is an outlier with all datapoints in the positive class.

First, we train the models including all the selected features, and evaluate the classification performance and the prioritization scores. Then, to quantify the contribution of diff-based information, we perform an ablation study in which we retrain the models after removing all commit-aware features.

The goal of the ablation study is to  evaluate whether the inclusion of commit-aware features produces statistically significant improvements in both classification and prioritization performance. We test the following hypotheses:
\begin{itemize}
    \item \(H_{01}\) (classification): the removal of diff-based features does not change classification performance;
    \item \(H_{02}\) (prioritization): removing diff-based features does not change the effectiveness of TCP.
\end{itemize}

\subsubsection{Statistical Analysis}

We apply the Wilcoxon signed-rank test for statistical significance and Cliff's delta for effect size.
As descriptive statistics, we report Q1, median (Q2), and Q3.

\section{Results and discussion}
\label{sec:results}

This section presents the experimental results, addressing the 3 research questions.

\subsection{RQ1: commit-aware feature effectiveness on test failure prediction}

The ablation study highlights the critical role of the diff features in the classification performance of both the XGBoost and the MLP models. When these features are removed,  performances degrade across all four projects, as shown in table \ref{tab:classificationPerformanceFolds}. Specifically, F1 and PR-AUC values drop from high-performance levels to near-zero values, indicating that the models lose their discriminative power without these specific inputs.

The analysis of the bootstrap quartiles reveals large differences in performance stability between the two experimental configurations. With diff-based features, the models exhibit a robust statistical profile, characterized by high median values and narrow interquartile ranges across the validation folds. This indicates that the predictive performance is not only elevated but also highly consistent, with minimal variance in the model's output. In  contrast, the removal of these features outlines a collapse of the distribution. In the ablation scenario, the entire quartile spread shifts  toward zero, and the upper bounds fail to reach meaningful classification thresholds. The absence of overlap between the distributions of the two scenarios confirms that the diff-based signals are not merely additive enhancements but are the foundation required for the models to achieve any discriminative capability.

Statistical analysis in table \ref{tab:statisticalAnalysis} confirms this impact. The Cliff's delta for both models and all classification metrics is 1.000, meaning a large effect size and demonstrating that the with-features configuration consistently outperforms the baseline across all folds. Furthermore, the Wilcoxon signed-rank test exhibits a p-value of 0.0625 for all metrics. Although the p-value does not cross the conventional 0.05 significance threshold, it corresponds to the smallest attainable p-value for a sample size of \(n = 4\) under this test, indicating the strongest evidence that could be obtained with the current experimental design.

\paragraph{Answer to RQ1:}
Based on the experimental results and statistical analysis, the hypothesis \(H_{01}\) can be rejected. The ablation of diff-based features leads to significant and systematic performance degradation in the classification task.

\begin{figure}[ht]
    \centering
    \includegraphics[width=1\linewidth]{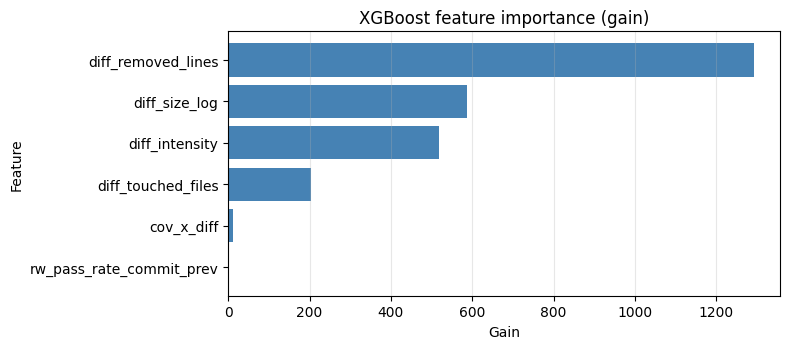}
    \caption{XGBoost feature importance on Lang test set}
    \label{fig:xgboostFeatureImportance}
\end{figure}

This finding is further supported by the feature importance chart shown in figure \ref{fig:xgboostFeatureImportance}: the diff-based features are the ones that most influence the decision-making process of the model.

\begin{table}[hbpt]
    \centering
    \caption{Classification report for both models on each fold. For each metric we report the Q1, Q2 and Q3 values. The number of bootstrap replications is 5000 per fold.}
    \begin{tabularx}{\columnwidth}{l c @{\hspace{1cm}} |  c @{\hspace{1cm}} c}
    \toprule
         \textbf{Metric} & \textbf{Diff features} & \textbf{XGBoost} & \textbf{MLP}  \\
         \midrule
         \multicolumn{4}{c}{\textbf{Lang}} \\
         \midrule
         
       \multirow{2}{*}{F1} & Yes &  (0.922, 0.939, 0.956) & (0.845, 0.870, 0.891) \\
                                & No & (0.000, 0.022, 0.039) & (0.126, 0.151, 0.180) \\
        \midrule
        \multirow{2}{*}{PR-AUC} & Yes & (0.857, 0.891, 0.923) & (0.898, 0.923, 0.951) \\
                                & No & (0.024, 0.028, 0.033) & (0.057, 0.074, 0.094) \\
        \midrule
        \multicolumn{4}{c}{\textbf{JacksonDatabind}} \\
        \midrule
        
       \multirow{2}{*}{F1} & Yes & (0.909, 0.941, 0.962) & (0.815, 0.857, 0.900) \\
                                & No & (0.018, 0.034, 0.051) & (0.000, 0.000, 0.000) \\
        \midrule
        \multirow{2}{*}{PR-AUC} & Yes & (0.964, 0.980, 0.996) & (1.000, 1.000, 1.000) \\
                                & No & (0.014, 0.018, 0.023) & (0.052, 0.072, 0.095) \\
        \midrule
         \multicolumn{4}{c}{\textbf{Jsoup}} \\
         \midrule
          
       \multirow{2}{*}{F1} & Yes & (0.863, 0.879, 0.895) & (0.291, 0.325, 0.359) \\
                                & No & (0.000, 0.000, 0.000) & (0.046, 0.061, 0.076) \\
        \midrule
        \multirow{2}{*}{PR-AUC} & Yes & (0.908, 0.928, 0.946) & (0.329, 0.364, 0.400) \\
                                & No & (0.069, 0.076, 0.083) & (0.050, 0.055, 0.060) \\
        \midrule
        \multicolumn{4}{c}{\textbf{Math}} \\
        \midrule
        \multirow{2}{*}{F1} & Yes & (0.882, 0.895, 0.908) & (0.743, 0.760, 0.778) \\
                                & No & (0.011, 0.015, 0.019) & (0.015, 0.019, 0.023) \\
        \midrule
        \multirow{2}{*}{PR-AUC} & Yes & (0.791, 0.817, 0.841) & (0.452, 0.477, 0.504) \\
                                & No & (0.009, 0.010, 0.011) & (0.007, 0.009, 0.010) \\
        \bottomrule
    \end{tabularx}
    \label{tab:classificationPerformanceFolds}
\end{table}

\subsection{RQ2: improved prediction effectiveness on fault detection}
The analysis of the prioritization quartiles reveals a distinct behavioral shift compared to the total collapse observed in the classification task, as shown in table \ref{tab:prioritizationPerformanceFolds}. The APFD Gain demonstrates a surprising resilience in the absence of diff-based features. This indicates that the models retain a good ability to order test suites effectively by leveraging residual signals, such as historical pass rates or coverage data. However, this resilience does not translate to the Speedup metric. Here, the removal of diff-based features causes a drastic degradation. This  dichotomy suggests that, while the models can achieve a generally good sorting quality without diff features, they lose the ability to identify the first failure immediately.

This dichotomy is statistically confirmed by table \ref{tab:statisticalAnalysis}: the Cliff’s $\delta$ for APFD gain indicates only a moderate effect size compared to the maximum separation seen in classification, whereas the Speedup metric reveals a large Mean $\delta$ Confidence Interval, confirming that, while general ordering is robust, the ability to rapidly detect failures  is strictly dependent on diff-based features.

\begin{table}[ht]
\centering
\caption{Cliff's \(\delta\) and Wilcoxon p-value and mean \(\delta\) CI (low-high interval) for XGBoost and MLP metrics on each fold prediction. Number of bootstrap samples: \(n=5000\); alternative hypothesis for Wilcoxon signed-rank test: \texttt{alternative='greater'}; number of pairs: \(p=4\). Mean \(\delta\) represents the per-fold difference between the two mean metric values from the with and without features scenario. }
\begin{tabularx}{\columnwidth}{c c | c @{\hspace{1.5cm}} c @{\hspace{1.5cm}} c  }
\toprule
\textbf{Model} & \textbf{Metric} & \textbf{Cliff \(\delta\)}& \textbf{p-value} & \textbf{Mean \(\delta\) CI (low-high)} \\
\midrule
\multirow{4}{*}{XGB} 
& F1                & 1.000 & 0.0625 & 0.877-0.912 \\
& PR-AUC                 & 1.000 & 0.0625 & 0.820-0.928 \\ 
& APFD gain & 0.500 & 0.0625 & 0.053-0.169 \\
& Speedup & 0.750 & 0.0625 & 258.750-1622.643 \\

\midrule
\multirow{4}{*}{MLP} 
                    & F1 & 1.000 & 0.0625 & 0.383-0.821 \\
                    & PR-AUC & 1.000 & 0.0625 & 0.388-0.889 \\
                    & APFD gain & 0.375 & 0.0625 & 0.140-0.509 \\
                    & Speedup & 0.500 & 0.0625 & 0.232-116.418 \\

\bottomrule
\end{tabularx}
\label{tab:statisticalAnalysis}
\end{table}

\paragraph{Answer to RQ2:}
The evidence regarding $H_{02}$ does not support rejection, as the prioritization performance gap is narrower. Although the removal of diff-based features severely impacts the time-to-first-failure (Speedup), the resilience of the APFD Gain, characterized by overlapping distributions, demonstrates that the overall efficiency of the test ordering is not entirely compromised.

\begin{table}[ht]
    \centering
    \caption{Prioritization metrics for both models on each fold. For each metric we report the Q1, Q2 and Q3 values. The number of bootstrap replications is 5000 per fold.}
    \begin{tabularx}{\columnwidth}{l c @{\hspace{1cm}} |  c @{\hspace{1cm}} c}
    \toprule
         \textbf{Metric} & \textbf{Diff features} & \textbf{XGBoost} & \textbf{MLP}  \\
         \midrule
          \multicolumn{4}{c}{\textbf{Lang}} \\
          \midrule
         \multirow{2}{*}{APFD Gain} & Yes & (0.506, 0.553, 0.599) & (0.501. 0.547, 0.594) \\
                                    & No & (0.453, 0.503, 0.552) & (0.477, 0.522, 0.571) \\
        \midrule
        \multirow{2}{*}{Speedup} & Yes & (809, 983, 1173) & (801, 975, 1168) \\
                                & No & (62, 79, 99) & (127, 161, 202)\\
        
        \midrule
        \multicolumn{4}{c}{\textbf{JacksonDatabind}} \\
          \midrule
           \multirow{2}{*}{APFD Gain} & Yes & (0.664, 0.711, 0.753) & (0.659, 0.709, 0.751) \\
                                    & No & (0.607, 0.654, 0.697)  & (0.596, 0.644, 0.685) \\
        \midrule
        \multirow{2}{*}{Speedup} & Yes & (504, 663, 745) & (568, 648, 728) \\
                                & No & (20, 25, 31) & (14, 16, 18) \\
        \midrule
      
         \multicolumn{4}{c}{\textbf{Jsoup}} \\
          \midrule
           \multirow{2}{*}{APFD Gain} & Yes & (0.541, 0.566, 0.589) & (0.440, 0.466, 0.492) \\
                                    & No & (0.326, 0.360, 0.392) & (-0.112, -0.073, -0.033) \\
        \midrule
        \multirow{2}{*}{Speedup} & Yes & (36, 43, 51) & (15, 19, 23) \\
                                & No & (6, 7, 8) & (6, 9, 14) \\
        \midrule
       
        \multicolumn{4}{c}{\textbf{Math}} \\
          \midrule
           \multirow{2}{*}{APFD Gain} & Yes & (0.459, 0.482, 0.505) & (0.460, 0.485, 0.510) \\
                                    & No & (0.334, 0.363, 0.392) & (0.395, 0.422, 0.450) \\
        \midrule
        \multirow{2}{*}{Speedup} & Yes & (1864, 2149, 2455)  & (3273, 4168, 5218) \\
                                & No & (88, 103, 121) & (78, 101, 126) \\
        
        \bottomrule
    \end{tabularx}
    \label{tab:prioritizationPerformanceFolds}
\end{table}

\subsection{RQ3: model comparison}
The comparative analysis of the bootstrap distributions definitively establishes XGBoost as the more robust and reliable model. In classification tasks (table \ref{tab:classificationPerformanceFolds}), XGBoost demonstrates superior stability, characterized by consistently narrower interquartile ranges and significantly higher median values across challenging projects. Regarding prioritization (table \ref{tab:prioritizationPerformanceFolds}), the distinction becomes nuanced: MLP occasionally achieves higher peak efficiency, as seen in the Math project, where its upper quartile Speedup reaches higher values. However, this peak performance is inconsistent, as the MLP fails to maintain this advantage in other contexts, dropping well below XGBoost.

\paragraph{Answer to RQ3:}
Whilst the neural network can sporadically outperform the ensemble method, XGBoost exhibits better stability. It offers the predictive consistency required for a reliable CI environment, avoiding the performance fluctuations inherent to the MLP architecture, while ensuring  operational advantages in terms of interpretability and implementation efficiency compared to an NN-based model.

\section{Conclusion}
\label{sec:conclusion}

We introduced a commit-aware, learning-based approach for Test Case Prioritization in Continuous Integration environments.
By integrating structural information extracted from version-control diffs with historical execution data and coverage metrics,
our method explicitly accounts for the characteristics of the current code change when prioritizing test execution.

Empirical evaluation on five Defects4J projects shows that commit-level signals play a central role in enabling accurate failure prediction.
The ablation study demonstrates that the removal of diff-based features leads to a systematic collapse of classification performance,
confirming that structural properties of code changes convey essential information about the likelihood of test failure.
Whilst prioritization metrics exhibit greater resilience to removal of commit-aware features, models equipped with these features consistently achieve faster fault detection
and more effective execution orders.
Across cross-project scenarios, tree-based models, particularly XGBoost, prove more robust and stable than neural models.

This study also highlights important limitations.
\paragraph{Internal threats:} data preprocessing, such as class rebalancing and selective imputation,
while necessary for effective learning, may move the setting away from a production-like CI pipeline.
\paragraph{Structural threats:} the approach relies on a heuristic mapping between production classes and test suites,
which enables scalable automation but may introduce noise when test–code relationships are complex.
\paragraph{External threats:} the evaluation is limited to Defects4J and results cannot be directly generalized to industrial pipelines with different scales and workflows.
In our study, models are trained offline in a static cross-project setting, whereas real CI environments evolve over time.

Despite these limitations, our results demonstrate that commit-aware TCP is both feasible and effective.
Future work will extend the evaluation to industrial datasets,
explore richer semantic representations of commits,
and investigate online and incremental learning strategies to improve adaptability to long-running CI pipelines.

\bibliographystyle{splncs04}
\bibliography{main}
\end{document}